\documentclass[11pt]{article}
\usepackage{latexsym}
\def\G{\Gamma}
\def\D{\Delta}
\def\d{\delta}
\def\a{\alpha}

\def\e{\epsilon}
\def\f{\phi}
\def\L{\Lambda}
\def\l{\lambda}
\def\g{\gamma}
\def\e{\epsilon}
\def\m{\mu}
\def\n{\nu}
\def\r{\rho}
\def\s{\sigma}
\def\S{\Sigma}
\def\O{\Omega}
\def\o{\omega}
\def\th{\theta}

\def\t{\tau}
\def\f{\phi}
\def\vf{\varphi}
\def\be{\begin{equation}}
\def\ee{\end{equation}}
\def\bea{\begin{eqnarray}}
\def\eea{\end{eqnarray}}

\def\ha{\frac{1}{2}}

\begin{document}
\centerline{\bf\Large Effective Actions for Regge Piecewise Flat}
\centerline{\bf\Large Quantum Gravity}

\bigskip
\begin{center}
{\bf A. Mikovi\'c} \footnote{Member of COPELABS and Mathematical Physics Group of the University of Lisbon.} \\
Departamento de Engenharia Informática e Sistemas de Informa\c{c}\~ao,  \\
Universidade Lus\'ofona de Humanidades e Tecnologias\\
Av. do Campo Grande, 376, 1749-024 Lisboa, Portugal\\
E-mail: amikovic@ulusofona.pt\\
\end{center}

\bigskip
\bigskip
\begin{quotation}
\noindent\small{We review the construction of the path integral and the corresponding effective action for the Regge formulation of General Relativity under the assumption that the short-distance structure of the spacetime is not a smooth 4-manifold, but a piecewise linear manifold based on a triangulation of a smooth 4-manifold. We point out that the exponentially damped 4-volume path-integral measure does not give a finite path integral, although it can be used for the construction of the perturbative effective action. We modify the 4-volume measure by multiplying it by an inverse power of the product of the edge-lengths such that the new measure gives a finite path integral while it retains all the nice features of the unmodified measure.}
\end{quotation}

\newpage
\section{Introduction}

Construction of a well-defined quantum gravity (QG) theory which has General Relativity (GR) as its classical limit has been an area of active research in the past 30 years. The difficulties with the Quantum Field Theory (QFT) quantization of GR led to the idea that the short-distance structure of the spacetime may be something different than a smooth 4-manifold. The most prominent approach of this type is String Theory \cite{sst}, which is based on the assumption that the short-distance structure of the spacetime is given by a loop manifold. Another prominent approach is Loop Quantum Gravity \cite{lqg}, where spin foams (2-complexes carrying the Lorentz group representations) are used to model the quantum spacetime. 

More recently, a new proposal involving a novel short-distance structure of the spacetime has appeard, which can be described as quantum gravity for piecewise flat spacetimes (PFQG) \cite{m, plqg}. In this paper we will review the PFQG approach in its simplest realization, which is based on the Regge formulation of GR, so that we will call it Regge PFQG. A PFQG theory is a quantum gravity theory which is based on a spacetime represented by a piecewise linear (PL) 4-manifold which is a triangulation of a smooth 4-manifold such that in each 4-simplex there is a flat metric. The simplest realization of such a PL manifold  is the Regge formulation of GR and the corresponding  QG theory is based on the Regge GR path integral \cite{R,RW}. 

Other examples of PFQG theories are the spin-foam models, see \cite{SF} for a review and references, and the spin-cube models \cite{scm}. However, in the spin-foam approach, as well as in the original Regge approach, the PL structure is considered as an auxilliary structure, which is introduced only for the purpose of constructing well-defined quantum transition amplitudes. It is expected that the PL structure will disapear after making the smooth-manifold limit. The same idea is used in the Causal Dynamical Triangulations approach \cite{cdt}, where the PL structure is removed by performing a sum over different triangulations.

If the spacetime is really a PL manifold based on a triangulation of a smooth manifold, then for a triangulation which has a large number of 4-simplices, the PL manifold will look like a smooth manifold at the scales much larger than the maximum edge length in the triangulation. Therefore, given a Regge path integral, one does not need to find its smooth-manifold limit (which is a difficult and still an unsolved problem) but instead it is sufficient to find a smooth-manifold approximation of the corresponding effective action. This is the same approach one uses when describing the dynamics of fluids, since a fluid consists of a large number of molecules, and at the length scales much larger than the inter-molecular distance, one can use a smooth vector field describing the average velocity of the molecules in a microscopic volume.

Another novelty of the PFQG approach is that a triangulation becomes a physical feature of the spacetime, so that at sufficiently small distances one would be able to see the effects of the PL structure. These distances are not necessarilly of the order of the Planck length ($10^{-35}\,m$), but can be larger. Note  that in the LHC experiments one can probe the distances down to $10^{-20}\, m$. Although no sign of quantum gravity effects was detected in the LHC experiments, still there are 15 of orders of magnitude to be explored.

In section 2 we review the standard Regge formalism for the Riemannian metrics and point out the problems with the corresponding Regge path integral. In section 3 we review the Regge calculus for pseudo-Riemannian metrics of the Minkowski signature, and define the corresponding Regge action. In section 4 we discuss the finitenness of the Regge path integral for a class of path-integral measures which allow the construction of the perturbative effective action. In section 5 we review in some detail the construction of the effective action, while in section 6  we explain how to construct the smooth-manifold approximation for the effective action. In section 7 we review the properties of the effective cosmological constant determined by the effective action, and discuss the implications for the cosmological constant problem. In section 8 we present our conclusions.

\section{Regge formulation of GR}

The Regge discretization of GR \cite{R,RW}, amounts to replacing the smooth spacetime manifold $M$ with a PL manifold $T(M)$ which corresponds to a triangulation of $M$. The metric on $T(M)$  is determined by the set of the edge lengths
\be\{ L_\e > 0 \,|\, \e \in T(M)\} \,,\label{sel}\ee
where $\e$ are the edges of $T(M)$. Although an $L_\e$ can be arbitrarilly small, we will exclude the zero edge lengths.

Given a set of the edge lengths (\ref{sel}), one would like to define a metric on the PL manifold $T(M)$  such that the PL metric in each 4-simplex $\s$ of $T(M)$ is flat and of the euclidean signature, i.e. $(+,+,+,+)$. This can be done by using the Cayley-Menger metric \cite{H}
\be G_{\m\n}(\s) = L_{0\m}^2 + L_{0\n}^2 - L_{\m\n}^2 \,,\label{cmm}\ee
where the five vertices of $\s$ are labeled as $0,1,2,3,4$ and $\m,\n = 1,2,3,4$. Although the CM metric is flat in a four-simplex, it is not dimensionless and hence it is not diffeomorphic to $g_{\m\n} = \d_{\m\n}$. This can be remedied by defining a new PL metric
\be g_{\m\n}(\s) = {G_{\m\n}(\s) \over \left(\det G(\s)\right)^{1/4}} \,,\label{eplm}\ee
which is a rescaled CM metric such that $\det\, g_{\m\n} = 1$, provided $\det\, G_{\m\n} > 0$.

Insuring the euclidean signature of a PL metric requires the following restrictions on the edge lengths:
\bea  \det\, G(\s) &>& 0 \,,\label{er1}\\
 \det\, G(\t) &>& 0 \,,\label{er2}\\
 \det\, G(\D) &>& 0 \,,\label{er3}\eea 
for every 4-simplex $\s$, every tetrahedron $\t$ and every triangle $\D$ of $T(M)$. The last inequality is equivalent to the triangular inequalities for the edge lengths of a triangle. These inequalities permit us to define the volumes of $n$-simplexes via the Cayley-Menger determinants \cite{CM}
\be \det\, G(\s_n ) = 2^n (n!)^2 V^2 (\s_n)\,,\quad n=2,3,4 \,. \ee

Note that for an arbitrary assignment of $L_\e$, the volumes $V_n$ can be positive, zero or imaginary. Just taking the strict triangular inequalities will ensure the positivity of the triangle areas, but then some of the higher volumes can be zero or imaginary. Hence all of the three inequalities must be imposed.

The Einstein-Hilbert (EH) action on $M$ is given by
\be S_{EH} = \int_M \sqrt{\det g}\, R(g) \, d^4 x  \,,\label{eh}\ee
where $R(g)$ is the scalar curvature associated to a metric $g$. On $T(M)$ the EH action becomes the Regge action 
\be S_R (L) = \sum_{\D\in T(M)} A_\D (L)\, \d_\D (L) \,,\label{ra}\ee
where $A_\D$ is the area of a triangle $\D$. The deficit angle $\d_\D$ is given by
\be \d_\D = 2\pi - \sum_{\s \supset \D} \th_\D^{(\s)} \,,\ee
where a dihedral angle $\th_\D^{(\s)}$ is defined as the angle between the 4-vector normals associated to the two tetrahedrons that share the triangle $\D$. Therefore
\be \sin \th_\D^{(\s)} = \frac{4}{3} {A_\D V_\s \over V_\t V_{\t'}} \,. \ee

Given the Regge action (\ref{ra}), the corresponding Euclidean path integral can be written as
\be Z_E = \int_{D} \prod_{\e=1}^{N} dL_\e \, \mu(L)\, e^{- S_R (L)/l_P^2} \,,\label{epi}\ee
where $D$ is the maximal subset of $({\bf R}_+ )^{N}$ consistent with the triangular inequalities\footnote{The number of the edge lengths $N$ can be made finite in the case of a non-compact manifold $M$ by labellenig only an appropriate compact subset of $T(M)$. For example, one can take $T(B_4)$, where $B_4$ is a four-ball in $M$.} . The path-integral measure $\m (L)$ is usually chosen as
\be \m (L) = \prod_{\e=1}^{N}  \left(L_\e\right)^\a  \,, \label{rm}\ee
where $\a$ is a constant, see \cite{H}, but other choices can be made, which we will discuss.

The immediate problem with the path-integral (\ref{epi}) and the measure (\ref{rm}) is that the finiteness of $Z_E$ is not guaranteed because $S_R(L)$ is not bounded from bellow, i.e. the scalar curvature can be unboundedly negative, and the measure (\ref{rm}) does not fall off sufficiently quickly for large $L_\e$ and negative $\a$. A simple way to remedy this is to complexify the Euclidean  path integral via
\be Z_{EC} = \int_{D} \prod_{\e=1}^{N}  dL_\e \, \mu(L)\, e^{i S_R (L)/l_P^2} \,.\label{cpi}\ee

This will be an absolutly  convergent integral for
\be \m(L) = \prod_{\e =1}^N  (l_0^2 + L_\e^2)^{-p} \,,\label{elm}\ee
where $p > 1/2$ and $l_0 > 0$, see \cite{scm}. However, the problem with (\ref{cpi}) is that it is not clear how to relate it to a path integral for the Minkowski signature metrics. 

\section{Lorentzian PL metric}

The problems with the euclidean path integrals (\ref{epi}) and (\ref{cpi}) can be avoided by using the Minkowski signature metric from the very beginning. In order to formulate the Regge action in the Lorentzian case,
we need to discuss certain aspects which are absent in the euclidean case.

The novelty in the Lorentzian case is that $L_\e^2$ can be positive or negative, so that $L_\e \in {\bf R}_+$ or $L_\e \in i\, {\bf R}_+$. Consequently we have to indicate in $T(M)$ which edges are space-like (S) and which edges are time-like (T). We will not use the light-like edges ($L_\e^2 = 0$). Although one can triangulate a pseudo-Riemannian manifold such that all the edges are spacelike \cite{m}, it is more natural to use the triangulations where we have both the spacelike and the timelike edges, see \cite{cdt}.

The CM metric is now given by the same expression as in the euclidean case (\ref{cmm}), while the physical PL metric is given by 
\be g_{\m\n}(\s) = {G_{\m\n}(\s) \over |\det\, G(\s)|^{1/4}} \quad,\ee
where the module of the determinant accounts for the fact that we now reqire $\det\, G(\s) < 0$.

In order to ensure the Minkowski signature of the PL metric we need to impose
\be \det\, G(\s) < 0 \quad, \ee
and
\be \det\, G(\t) \ne 0 \quad,\quad \det\, G(\D) \ne 0 \quad, \ee
for any $\s$, $\t$ and $\D$ in $T(M)$. This is analogous to the first restriction in the euclidean case (\ref{er1}). The analogs of the second (\ref{er2}) and the third restriction (\ref{er3}) are weaker, since the signatures of $\det\, G(\t)$ and $\det\, G(\D)$ are not fixed in the Minkowski case. Namely, $ \det\, G(\t) > 0$ if $\t$ belongs to a euclidean hyper-plane of $g_{\m\n}(\s)$, while $\det\, G(\t) < 0$ if $\t$ belongs to a Minkowski hyper-plane. Also $\det\, G(\D) > 0$ if $\D$ belongs to a euclidean 
plane while $\det\, G(\D) < 0$ if $\D$ belongs to a Minkowski plane.

The volumes of $n$-simplexes can be defined as
\be (V_n)^2 =  {|\det\, G_n |\over 2^n (n!)^2 } > 0 \,,\quad n = 2,3,4 \,,\ee
so that $V_n > 0$. Note that in the $n=1$ case we should distinguish between the labels $L_\e \in {\bf C}$ and their 1-volumes as $|L_\e | > 0$. We will also use an equivalent labeling $L_\e \to |L_\e|$ with an indication $S$ for $L_\e \in {\bf R}_+$ or $T$ for $L_\e \in i\,{\bf R}_+$ . In order to avoid pathological zero volumes, we will require
\be \det\, G(\t) \ne 0 \,, \quad \det\, G(\D) \ne 0 \,,\ee
for all tetrahedrons and triangles in $T(M)$.

Given that the edge lengths can take real or imaginary values in a Minkowski space, this implies that the angles between the vectors can be real or complex. Let us consider the angles in a Minkowski plane. Such angles can be defined as
\be \cos\a = {\vec u \cdot \vec v \over ||\vec u||\,||\vec v||}\,,\quad \sin \a = \sqrt{1 -\cos^2 \a}\,, \quad \a\in {\bf C}\,,\label{mia}\ee
where $\vec u = (u_1, u_0)$, $\vec u \cdot \vec v = u_1 v_1 - u_0 v_0 $ and $||\vec u || =\sqrt{ \vec u \cdot \vec u}$. 

Consider two spacelike vectors $\vec u = (1,0)$ and $\vec v = (\cosh a, \sinh a )$, $a\in \bf R$. Since $||\vec u || = ||\vec v || = 1$ then
\be \cos\a = \cosh a \,,\quad \sin\a = i \sinh a \,\Rightarrow \,\a =  i\,a \,.\ee

In the case of a spacelike vector $\vec u = (1,0)$ and a timelike vector $\vec v = (\sinh a, \cosh a )$, we have
$||\vec u || = 1$ and $||\vec v || = i$, so that
\be \cos\a = -i\sinh a \,,\quad \sin\a =  \cosh a \,\Rightarrow \,\a = \frac{\pi}{2} - i\,a \,.\ee

And if we have two timelike vectors $\vec u = (0,1)$ and $\vec v = (\sinh a, \cosh a )$, then
\be \cos\a = \cosh a \,,\quad \sin\a = i \sinh a \,\Rightarrow \,\a =  i\,a \,.\ee

The definition (\ref{mia}) then implies that the sum of the angles between two intersecting lines in a Minkowski plane is $2\pi$.

In order to define the dihedral angles in the Minkowski case we will introduce 
\be (v_n)^2 =  {\det\, G_n \over 2^n (n!)^2 }  \,,\quad n = 2,3,4 \,,\ee
so that $v_n = V_n$  for $\det\,G_n > 0$ or $v_n = i\, V_n$ for $\det\,G_n < 0$. In the $n=1$ case we have $v_\e = L_\e$ for a spacelike edge or $v_\e = iL_\e$ for a timelike edge where $L_\e > 0$. Then the angle between two edges in a triangle is given by
\be \sin \a_\pi^{(\D)} = {2\,v_\D \over v_\e \,v_{\e'}}\,,\label{lda}\ee
where $\pi$ the common point (known as the hinge).

The dihedral angle between two triangles sharing an edge in a tetrahedron is given by
\be  \sin \f_\e^{(\t)} =\frac{3}{2}\, { v_\e \,v_\t \over v_\D \,v_{\D'}}\,,\label{ldb}\ee
while the dihedral angle between two tetrahedrons sharing a triangle in a four-simplex is given by
\be \sin \th_\D^{(\s)} = \frac{4}{3}\,{ v_\D \, v_\s \over v_\t \,v_{\t'}}\quad.\label{ldc}\ee

The formulas (\ref{lda}),(\ref{ldb}) and (\ref{ldc}) are generalizations the corresponding euclidean formulas such that $V_n \to v_n$, and the novelty in the Minkowski case is that $\sin \th$ is not restricted to the interval $[-1,1]$, but $\sin \th \in \bf R$ or $\sin \th \in i{\bf R}$. This also means that the Minkowski dihedral angles can take the complex values.

In the case of a dihedral angle $\th_\D^{(\s)}$ there are two possibilities. If the triangle $\D$ is in a Minkowski (ST) plane, then $\th$ will be an angle in an orthogonal Euclidean (SS) plane, so that
$\sin\th = \sin a$. If $\D$ is in an SS plane, then $\th$ will be in an orthogonal ST plane, so that $\sin\th = \cosh a$ or $\sin\th =i\sinh a$. 

The deficit angle will then take the following values:
\be\d_\D = 2\pi - \sum_{\s\supset\D} \th_\D^{(\s)} \in {\bf R} \,,\ee
when $\D$ is an ST triangle, while
\be\d_\D = 2\pi - \sum_{\s\supset\D} \th_\D^{(\s)} \in  \frac{\pi}{2}{\bf Z} + i{\bf R} \,,\ee
when $\D$ is an SS triangle. Note that for an SS triangle the triangle inequalities are valid, while for an ST triangle they do not apply.

The appearance of the complex values for the deficit angles in the Minkowski signature case raises the question of how to generalize the euclidean Regge action such that the new action is real. A proposal for a lorentzian Regge action was given in \cite{cdt}
\be   S_R = \sum_{\D \in SS} A_\D \, \frac{1}{i}\,\d_\D + \sum_{\D \in ST} A_\D \,\d_\D  \,.\label{ma} \ee
However, the problem with this definition is that a priori $S_R \in {\bf R} + i \frac{\pi}{2}{\bf Z}$, so that one has to verify for a given triangulation that $Im\, S_R = 0$.

In order to avoid this difficulty, we will take
\be  S_R = Re\left(\sum_{\D \in SS} A_\D \, \frac{1}{i}\,\d_\D \right) + \sum_{\D \in ST} A_\D \,\d_\D  \,.\label{mra} \ee
This definition can be justified by the fact that the authors of \cite{cdt} have verified that $Im\,\tilde S_R = 0$ for a special class of triangulations, which are physically relevant, and they are called the casual triangulations.

\section{GR path integral}

Given the definition of the Regge action in the Lorentzian metric case (\ref{mra}), we can write
\be Z (T(M), \m) = \int_D \prod_{\e = 1}^N dL_\e \,\m(L)\,e^{iS_R(L)/l_P^2} \,,\label{rpi}\ee
where $D$ is the maximal subset of ${\bf R}_+^N$ consistent with a choice of the spacelike and the timelike edges. 

The convergence of the integral (\ref{rpi})
is of the fundamental importance for the Regge PFQG theory. Also, in order to have an effective action with the correct classical limit we have to use a PI measure which satisfies the following criterion 
\be\ln \m (\l\, L_1, \cdots, \l L_N)\approx  O(\l^a)\,,\quad a \ge 2 \,,  \label{asm}\ee
for $\l\to \infty$, see \cite{m}. 

A choice for $\m(L)$ can be made such that it obeys (\ref{asm}) and it insures the diffeomorphism invariance of the  smooth-spacetime effective  action \cite{m}. It is given by
\be \m(L) = \exp\left(-V_4 (L)/L_0^4 \right) \,,\label{plm}\ee
where $V_4$ is the volume of $T(M)$ and $L_0$ is a new parameter in the theory, see \cite{m}. 

Although  (\ref{plm}) is an exponentially damped measure when $L_\e \to \infty$, the convergence of the path integral  is not guaranteed beacuse of the existence of the degenerate configurations where some or all of the $L_\e \to\infty$ while $V_4 \to 0$. If we denote the union of such regions as $D_0$ and let $D' = D \setminus D_0$, then
\be |Z| \le \int_{D} \prod_\e dL_\e \, e^{-V_4(L)/L_0^4} \approx \int_{D_0} \prod_\e dL_\e +  \int_{D'} \prod_\e dL_\e \, e^{-V_4(L)/L_0^4} \,. \ee
The integral over $D_0$ is clearly divergent, so that we cannot prove the absolute convergence of $Z$. This does not mean that $Z$ is divergent, but it is an indication that we have to check the convergence of $Z$ in some other way.

By using the spherical coordinates in ${\bf R}_+^N$, the path integral $Z$ can be written as
\be Z = c_N \,\int_{0}^\infty L^{N-1}\, dL \int_{\O_N} d^{N-1}\theta J_N(\theta)\, e^{-L^4 v(\theta)/L_0^4 +i L^2 s(\theta)/l_P^2}\,,  \ee
where 
$$L^2 = \sum_{\e =1}^N L_\e^2 \,,$$
and
$$  \theta = (\th_1, \th_2, \cdots, \th_{N-1}) \,,$$
are the angles corresponding to the points of ${\bf R}_+^N$. $c_N$ is a constant and $J_N (\th)$ denotes the angular dependence of the Jacobian.

If we integrate first the $L$ variable, we obtain
$$Z = c_N \int_{\O} d^{N-1}\theta J_N(\theta)\, \int_{0}^\infty L^{N-1} \, e^{-v(\th)L^4 /L_0^4 +is(\theta)L^2/l_P^2}\, dL $$
\be = \int_{\O} d^{N-1}\theta J_N(\theta)\, F(v(\theta),s(\theta))\,, \label{angi}\ee
where $\O$ is the angular region corresponding to the path-integral integration region $D\subset {\bf R}_+^N$.

The function $F(v,s)$ has the following asymptotic properties. For $s\to 0$ we have
$$ F(v,s) \approx c_N \int_{0}^\infty L^{N-1} \, e^{-vL^4/L_0^4 }\, dL $$
\be = c_N v^{-N/4}\,\int_{0}^\infty \xi^{N-1} \, e^{-\xi^4 /L_0^4 }\, d\xi = c'_N  \, v^{-N/4} \,.\label{sa}\ee

Similarly, for $v \to 0$ we obtain
\be F(v,s) \approx c_N \int_{0}^\infty L^{N-1} \, e^{-sL^2/l_P^2 }\, dL = c''_N\,s^{-N/2} \,.\label{va}\ee

Since $\O$ is a compact set, the angular integral (\ref{angi}) will diverge if $F$ is not bounded in $\O$. From (\ref{sa}) and (\ref{va}) we can see that $F$ is not bounded when $v\approx 0$ and $s\approx 0$. Note that $v\approx 0$ corresponds to $V_4(L) \approx 0$, and $s\approx 0$ corresponds to $S_R(L) \approx 0$. Hence the integral $Z$ with the measure (\ref{plm})  is divergent.

Let us now consider the following class of PI measures
\be \m_p(L) = \prod_\e (l_0^2 + L_\e^2)^{-p} \, \exp\left(-V_4 (L)/L_0^4 \right) \,,\label{mup}\ee
such that $p > 1/2$ and $l_0 > 0$. This is a product of the measures (\ref{elm}) and (\ref{plm}) and we have
$$ |Z| \le \int_{D} \prod_\e {dL_\e \over (l_0^2 + L_\e^2)^{p}}\, e^{-V_4(L)/L_0^4} \approx \int_{D_0} \prod_\e {dL_\e \over(l_0^2 + L_\e^2)^{p}} + \int_{D'} d^N L \,\m_p(L) \,.$$

The first integral is convergent for $p > 1/2$, while the second integral is also convergent, and this can be shown by using the $N$-dimensional spherical coordinates in $D'$. We then have
$$|Z| \le  I_2 = C + c_N\int_{L'}^\infty L^{N-1}\, dL \int_{\O'} d^{N-1}\theta J_N(\theta)\, {e^{-L^4 v(\theta)/L_0^4}\over L^{2Np}}\,  \,,$$
where $L' \gg l_0$. We can write
$$I_2 \approx C + c_N\int_{\O'} d^{N-1}\theta J_N(\theta)\, (v(\theta))^{N(p-1/2)/2}\int_{\xi'}^\infty \xi^{N-1} \,e^{-\xi^4 /L_0^4}\, d\xi $$
$$ = C + c'_N\int_{\O'} d^{N-1}\theta J_N(\theta)\, (v(\theta))^{\frac{N}{2}\left(p-\frac{1}{2}\right)}\,.$$

The angular integral in $I_2$ is finite, because it is an integral of a bounded function over a compact region. Hence $Z$ with the measure (\ref{mup}) is absolutely convergent. This measure also satisfies the criterion (\ref{asm}) for the construction of the effective action, so that it can be used to define a PFQG theory. 

One wonders  whether the $p=0$ mesure could be still used, in spite of the fact that $Z$ is divergent. We will see in the next section that  the construction of the perturbative effective action does not depend on the finiteness of $Z$. However, if one wants to construct a non-perturbative effective action, then the finiteness of $Z$ is important.

\section{Effective action}

The effective action is an important tool for understanding the semiclassical properties of a quantum theory, because it is an action which gives the EOM for quantum corrected classical trajectories. It was first introduced for QFTs by using the path-integral formulation, and one can also use an effective action to understand the non-perturbative features, because an effective action is a generating functional for the one-particle irreducible correlation functions. 

For the purposes of constructing an effective action for a Regge PFQG, we need to define the path integral for the case of a manifold $M = \S\times [a,b]$, where $\S$ is a 3-manifold. Let $[a,b] = [0,nt]$, $n\in \bf N$ and $t>0$. We will use a time-ordered triangulation, which is also known as a causal triangulation \cite{cdt}. Let
\be T(M) = \cup_{k = 0}^{n-1} \, \tilde T_k \left(\S \times [k,k+1]\right) \,, \label{ct}\ee
where $\tilde T_k$ is a triangulation of a slab $\S\times [k,k+1]$ such that 
$$\partial \tilde T_k = T_k (\S) \cup T_{k+1}(\S)$$ 
and $T_k$ are triangulations of $\S$.  We then choose $v_\e = L_\e$ for $\e \in T_k (\S)$ and $v_\e = i L_\e$ for $\e\in \tilde T_k \setminus (T_k \cup T_{k+1})$.

We will determine the semiclassical properties of a Regge PFQG by using a perturbative effective action. It can be constructed by solving the effective action equation in the semiclassical limit $L_\epsilon \gg l_P = \sqrt{G_N  \hbar}$.

Let us recall first the effective action definition from quantum field theory. Let $\phi$ be a real scalar field on $M$ and let
$$ S (\phi) = \frac{1}{2}\int_M d^4 x \sqrt{|g|}\left[g^{\mu\nu}\,\partial_\mu \phi \,\partial_\nu \phi -  m^2\,  \phi^2 - \lambda\,\phi^4 \right]\,,$$
be a flat-spacetime action. The effective action $\Gamma (\phi)$  can be determined from the following integro-differential equation
\be e^{i\Gamma(\phi)/\hbar}=\int {\cal D} h \exp\left[{ i\over\hbar} S(\phi + h) -{i\over\hbar} \int_M d^4 x\,{\delta\Gamma\over\delta\phi (x)}h(x)\right]\,, \label{qftea}\ee
see \cite{n,k}.

The equation (\ref{qftea}) follows from the definitions
$$ Z(J) = \int {\cal D}\vf \, e^{\frac{i}{\hbar}[S(\vf) + \int_M d^4x J(x)\vf(x) ]} = e^{\frac{i}{\hbar} W(J)}$$
and
$$ \G(\f) = W(J) - \int_M d^4x J(x)\f(x)\,, $$
so that
$$   e^{\frac{i}{\hbar}\G(\f)} = \int {\cal D}\vf \, e^{\frac{i}{\hbar}[S(\vf) + \int_M d^4x J(x)(\vf(x) - \f(x)) ]}\,. $$

By changing the integration variable $\vf(x)$ to $h(x) = \vf(x) -\f(x)$, we obtain the equation (\ref{qftea}). Note that the limits of integration for the variable $h$ are the same as those for $\vf$, since
$$ \vf \in (-\infty, \infty) \rightarrow h = \vf - \f \in (-\infty, \infty) \,.$$
However, in the case when $\vf \in I \subset {\bf R}$, then the integration limits of the variable $h$ become $\f$-dependent. For example
$$\vf \in (0, \infty) \rightarrow h = \vf - \f \in (-\f, \infty) \,.$$
This happens in the case of the Regge action, since then $\f(x) \sim L_\e > 0$ and $h(x) \sim l_\e$, see the equation (\ref{eae}).

The equation (\ref{qftea}) can be solved perturbatively as
$$ \G(\f) = S(\f) + \hbar \G_1 (\f) + \hbar^2 \G_2 (\f) + \cdots \,,$$
and a perturbative solution will be a complex-valued function. A real effective action is obtained by using the Wick rotation. This is done by solving first the EA equation in the Euclidean spacetime
\be e^{-\Gamma_E (\phi)/\hbar}=\int {\cal D} h \exp\left[-{ 1\over\hbar} S_E (\phi + h) +{1\over\hbar} \int_M d^4 x\,{\delta\Gamma_E \over\delta\phi (x)}h(x)\right]\,,\ee
so that all the solutions are real. Then $x_0 = -it$ is  inserted into a solution $\Gamma_E (\phi)$, where $(x_0, x_k)$ are the spacetime coordinates, and one takes
$$\G (\phi) = - \G_E (\phi)|_{x_0 = -it} \,.$$

However, the Wick rotation cannot be used in quantum gravity, since in many problems of interest, introducing a flat background metric does not make sense. One way to resolve this difficulty is to use the fact that the Wick rotation in QFT is equivalent to 
\be \Gamma(\phi) \to Re\,\Gamma(\phi) + Im\,\Gamma(\phi)\,,\ee
where $\G(\f)$ is a perturbative solution of (\ref{qftea}), see \cite{m,mv}. This prescription is convenient for quantum gravity because it does not involve a background metric, nor a system of coordinates.

In the case of Regge quantum gravity without matter, the effective action equation is given by 
\be e^{i{\Gamma}(L)/l_P^2} = \int_{D(L)} d^N  l \, \mu (L +l) e^{iS_{Rc} (L+l)/l_P^2 - i\sum_{\epsilon=1}^N {\Gamma}'_\epsilon (L)l_\epsilon /l_P^2 } \,,\label{eae}\ee
where $S_{Rc}$ is the Regge action plus the cosmological constant term and $D(L)$ is a subset of ${\bf R}^N$ obtained by translating $D$ by a vector $-L$ \cite{m}. Note that $D(L) \subset  [-L_1, \infty)\times\cdots\times[-L_N, \infty)$. 

We will look for a semiclassical solution
$$ \Gamma (L) = S_{Rc} (L) + l_P^2 \Gamma_1 (L) + l_P^4 \Gamma_2 (L) + \cdots \,,$$
where $L_\epsilon \gg l_P$ and
$$| \Gamma_n (L) | \gg l_P^2 |\Gamma_{n+1}(L) |\,. $$

When $L_\epsilon \to \infty$, then $D(L) \to {\bf R}^N$ and
\be e^{i{\Gamma}(L)/l_P^2} \approx \int_{{\bf R}^N} d^N l \, \mu (L +l) e^{iS_{Rc} (L+l)/l_P^2 - i\sum_{\epsilon=1}^E {\Gamma}'_\epsilon (L)l_\epsilon /l_P^2 } \,.\label{llea}\ee

Actually, one can use the equation (\ref{llea}) to determine $\G(L)$ for large $L$ when $\m$ falls off sufficiently quickly \cite{m}. The reason is that 
$$D (L) \approx [-L_1,\infty)\times\cdots\times[-L_N,\infty)\,,$$
for $L_\e \to \infty$, so that the relevant behaviour is captured by the following one-dimensional integral
$$\int_{-L}^\infty dx \, e^{-zx^2/l_P^2 - wx} = \sqrt{\pi}\,l_P \exp{\Big [}-\frac{1}{2} \log z + l_P^2 {w^2 \over 4z} $$
$$+l_P{ e^{-z\bar L^2 /l_P^2}\over 2\sqrt{\pi z}\bar L}\left(1 + O(l_P^2 /z\bar L^2)\right) {\Big ]} \,,$$
where $\bar L = L + l_P^2 {w\over 2z}$ and $Re\, z =- (\log\mu )'' $. The non-analytic terms in $\hbar$ will be absent if
\be \lim_{L\to\infty}e^{-z\bar L^2 /l_P^2} = 0 \Leftrightarrow (\log\m)'' < 0 \,\,\textrm{for}\,\, L \to\infty\,. \label{tmr} \ee

Hence the perturbative solution exists for the exponentially damped measures, and the toy model requirement (\ref{tmr}) becomes the criterion (\ref{asm}) when we have more than one $L_\e$.

For $D(L) ={\bf  R}^N$ and $\mu(L)$ a constant, the perturbative solution is given by the EA diagrams
$$ \Gamma_1 ={i\over 2}Tr \log S''_{Rc}  \,,\quad \Gamma_2 = \langle S_3^2 G^3 \rangle + \langle S_4 G^2 \rangle\,, $$
and
$$ \Gamma_3 = \langle S_3^4 G^6 \rangle + \langle S_3^2 S_4 G^5  \rangle + \langle S_3 S_5 G^4  \rangle + \langle S_4^2 G^4  \rangle + \langle S_6 G^3 \rangle \,, \,... $$
where $G = i(S_{Rc}'')^{-1}$ is the propagator and $S_n = iS_{Rc}^{(n)}/n!$ for $n > 2$, are the vertex weights, see \cite{k,m}. The contractions $\langle X \cdots Y \rangle$ are the sums over the repeated DOF indices
$$\langle X\cdots Y\rangle =  \sum_{k,...,l} X_{k...l} \cdots Y_{k...l}\quad. $$

When $\mu(L)$ is not a constant, then the perturbative solution is given by
$$ \Gamma (L) = \bar S_{Rc} (L) + l_P^2 \bar\Gamma_1 (L) + l_P^4 \bar \Gamma_2 (L) + \cdots \,,$$ 
where 
$$\bar S_{Rc} = S_{Rc} - il_P^2 \log\mu \,,$$
while $\bar\Gamma_n$ is given by the sum of $n$-loop EA diagrams with $\bar G$ propagators and $\bar S_n$ vertex weights \cite{m}.

Therefore
$$ \Gamma_1 = -i\log\mu + {i\over 2}Tr \log S''_{Rc} $$
$$ \Gamma_2 = \langle S_3^2 G^3 \rangle + \langle S_4 G^2 \rangle + Res [l_P^{-4} Tr\log\bar G ]\,, $$
$$ \Gamma_3 = \langle S_3^4 G^6 \rangle + \cdots + \langle S_6 G^3 \rangle + Res [l_P^{-6} Tr\log\bar G ] + Res[l_P^{-6}\langle \bar S_3^2 \bar G^3 \rangle] + Res[l_P^{-6}\langle \bar S_4 \bar G^2 \rangle]\,,  $$
see \cite{m}.

Since the PI measure $\m(L)$ has to vanish exponentially for large edge lengths, a natural choice is the measure (\ref{plm}). This measure satisfies the criterion (\ref{asm}) for the classical limit, since 
\be\log\mu(L) = O\left( (L /L_0)^{4} \right) \,,\label{ma}\ee
where the notation $f(x_1,...,x_n) = O(x^\a)$ means that 
$$f(\l x_1,...,\l x_n) = O(\l^\a)$$ 
for $\l\to\infty$.

Then for $L_\epsilon > L_c$ and
\be L_0 > \sqrt{l_P\, L_c}\,,\label{pecc}\ee  
where $L_c^{-2} = \L_c$, we get the following large-$L$ asymptotics \cite{mv,mvp}
\be\Gamma_1 (L) = O(L^4/L_0^4) + \log O(L^2/L_c^2) + \log\theta(L) + O(L_c^2 /L^2) \label{go}\ee
and
\be  \Gamma_{n+1}(L) = O\left((L_c^2 /L^4)^{n}\right) + L_{0c}^{-2n} O\left((L_c^2 /L^2)\right)\,, \label{gn}\ee
where $L_{0c} = L_0^2 /L_c $.

Note that the construction of the perturbative EA only requires the criterion (\ref{asm}), while the finitennes of $Z$ is not necessary. However, if we want to construct exact (non-perturbative) solutions of the EA equation, this requires that the integral in (\ref{eae}) is finite. If the PI mesure is such that $Z$ is absolutely convergent, than the integral in (\ref{eae}) will be also absolutely convergent.

The modified PI measure (\ref{mup}) gives an absolutely convergent $Z$, and satisfies the criterion (\ref{ma}), so that we can use it to define a Regge PFQG.

\section{The smooth-manifold approximation}

In order to understand the effects of the Regge effective action for a smooth spacetime, we need to see how to approximate the Regge effective action with a QFT effective action. 

Let $T(M)$ has a large number of the edges ($N \gg 1$) and let the variation of the edge lengths from each triangle to its neighbour be small. Given a function $f(L)$, we would like to approximate it with a functional of a smooth metric on $M$. The smooth limit can be defined as the limit $N \to\infty$ and $L_\e  \to 0$ such that
$$g_{\m\n}^{(\s)}(L) \to g_{\m\n}(x) \,,$$
where $x$ are the coordinates of a point inside the 4-symplex $\s$ and the partial derivatives  of $g_{\m\n}(x)$ are continious on $M$ up to order $n \ge 2$ .

In the case of the Regge action, for large $N$ and a small local variation of the edge lengths, there is a smooth metric on $M$ such that 
\be S_{R} (L) \approx {1\over 2} \int_M d^4 x \sqrt{|g|} \, R(g) \,.\ee
We also have 
\be \Lambda_c V_4 (L) \approx \Lambda_c \int_M d^4 x \sqrt{|g|} = \L_c \,V_M \,, \ee
where $|g| = |\det \,g|$. These are the standard formulas of the Regge calculus and they nicely illustrate how functions $f(L)$ on the PL manifold $T(M)$ can be approximated by functionals of a smooth (differentiable) metric $g$ on $M$ for $N\to\infty$.

Similarly, the effective action $\G(L)$ can be approximated by a QFT effective action $\tilde\G (g)$, where $g$ is a smooth metric on $M$. In the region where $L_\e \ge L_K$ such that $L_K \gg l_P$, the following approximation is valid 
\be Tr \log S_R''(L) \approx \int_M d^4 x \sqrt{|g|} \left( a R^2 + b R_{\mu\nu}R^{\mu\nu}  + \cdots \right)\ln{K\over k_0}\label{ola}\ee
where $a$ and $b$ are numerical constants, while $k_0$ is an arbitrary constant such that $k_0 \ll K$. The $\cdots$ indicate some additional terms which may be present in the QFT effective action, like 
$$ c\, R \log\left({\Box \over m_0^2}\right) R + d\, R_{\m\n} \log\left({\Box \over m_0^2}\right) R^{\m\n}  \,,$$
see \cite{BV}, where $\Box = g^{\m\n}\nabla_\m \nabla_\n$ and $m_0 = 1/k_0$.

The formula (\ref{ola}) follows from the fact that a PL function on a lattice with a cell size $L_K$ can be written as a Fourier integral over a compact region $|q|\le \pi/L_K$ where $q$ is the wave vector\footnote{This region is known as the first Brillouin zone.}. Hence the PL trace-log term can be approximated by the one-loop QFT effective action for GR by using a momentum cutoff $K = 2\pi\hbar/L_K$. That is how one obtains the approximation (\ref{ola}).

Beside the standard trace-log term, the first-order effective action  will also contain the $\log\m$ term, so that
\be\G_1(L)  =Tr(\log S''_R(L))  + {V_4(L)\over L_0^4} - p\sum_\e \ln \left(1+{L^2_\e \over l_0^2}\right)  \,,\label{pgfoea}\ee 
where we have discarded the constant term $-p N \ln l_0^2$. 

The last term in (\ref{pgfoea}) cannot be expressed as a functional of a metric in the smooth-manifold approximation. However, one can argue that it becomes negligibly small when $N\to\infty$ and $L_\e\to 0$. Let $L_\e \approx l_0 /N$, then 
$$ \sum_\e \ln \left(1+{L^2_\e \over l_0^2}\right) \approx N \ln \left(1 + {1\over N^2}\right) \approx {1\over N} \,.$$

Hence the measure correction term will not affect the smooth-manifold approximation. However, when $L_\e \to \infty$, the measure correction term will be subdominant with respect to the $V_4(L)$ term, since 
$$ V_4(L) = O(L^4) \,,\quad \log \m_p - \log\m_0 = O(\ln L) \,, $$
so that $\m_p$ will have the same semiclassical properties as $\m_0$.

\section{Effective cosmological constant}

The asymptotics (\ref{go}) and (\ref{gn})  imply that the series 
$$\G(L) = \sum_{n\ge 0} ( l_P)^{2n}\Gamma_n (L)$$
is semiclassical (SC) for $L_\epsilon \gg l_P$ and  $ L_0 \gg \sqrt{l_P\, L_c}$.

Let $\Gamma \to \Gamma /G_N$ so that $ S_{eff} = ( Re\,\Gamma + Im\,\Gamma) /G_N$.
The effective action is then given by
$$ S_{eff} = S_{Rc}  + {l_P^2 \over L_0^4} V_4 + {l_P^2 \over 2} Tr\log S''_{Rc}  - p\,l_P^2 \sum_\e \ln (1 + L_\e^2 /l_0^2 )  + O(l_P^4)\,, $$
for $L_\e \gg l_P$. For the sake of simplicity, we will consider only the $p=0$ case, since $p > 0$ cases  are essentially the same. Hence  the $O(\hbar)$, or the one-loop, cosmological constant (CC) for pure gravity is given by
\be \Lambda = \L_c + {l_P^2 \over  L_0^4} = \L_c + \L_{qg} \,.\label{qgcc}\ee

One can show that the one-loop cosmological constant is exact because there are no $O(L^4)$ terms beyond the one-loop order \cite{mv,mvp}. This is a consequence of the large-$L$ asymptotics
$$ \log \bar S''_{Rc}(L) = \log O(L^2 /\bar L_c^2) + \log\theta(L) +O(\bar L_c^2 /L^2) $$
$$\bar\G_{n+1} (L) = O\left((\bar L_c^2 /L^4)^n \right) \,,$$
where $\bar L_c^2 = L_c^2 \left[ 1 + il_P^2 ( L_c^2 /L_0^4 )\right]^{-1/2}$.

Hence the one-loop formula (\ref{qgcc}) is exact in the case of pure gravity.
If $\L_c =0$, the observed value of $\Lambda$ is obtained for $L_0 \approx 10^{-5} m$ so that $ l_P^2 \Lambda \approx 10^{-122}$ \cite{m}. Note that $L_0 \approx 10^{-5} m$ is consistent with the requirement that $L_0 \gg l_P$, which replaces the SC condition $L_0 \gg \sqrt{L_c l_P}$ when $\L_c = 0$. 

The formula (\ref{qgcc}) is intriguing but unrealistic, since there is matter in the universe. In order to obtain a realistic expression for the effective CC, we need to study the EA equation with matter. 
 
The effect of the matter on the CC can be studied by introducing a scalar field on $M$
\be S_m (g,\phi) = \frac{1}{2}\int_M d^4 x \sqrt{|g|}\left[g^{\mu\nu}\,\partial_\mu \phi \,\partial_\nu \phi -  U(\phi) \right]\,,\label{scfa}\ee
where $U={1\over 2}\o^2  \phi^2 + \lambda\,\phi^4$, where $\o = m/\hbar$.
 
On a PL manifold $T(M)$ the action (\ref{scfa}) becomes
$$ S_{m} = \frac{1}{2}\sum_\sigma V_\sigma (L)  \sum_{k,l} g^{kl}_\sigma (L)\,  \phi'_k \, \phi'_l - \frac{1}{2}\sum_p V_p^* (L)\, U( \phi_p) \,,$$
where $\phi'_k = (\phi_k -\phi_0 ) /L_{0k}$ and $k,l,0$ are vertices in a 4-simplex $\s$, $p$ labels the vertices of $T(M)$ and $V^*$ is the volume of the dual cell. Then the total classical action of gravity plus matter on $T(M)$ is given by
$$ S(L,\phi) ={1\over G_N} S_{Rc}(L) + S_m (L,\phi) \,. $$

The corresponding EA equation is given by
\bea e^{i\Gamma (L,\phi)/l_P^2} = \int_{D_E (L)} d^E l \,\int_{{\bf R}^V}\,d^V \chi &\exp & \Big{[} i \bar S (L+l, \phi + \chi)/l_P^2 
-i\sum_\epsilon \frac{\partial\Gamma}{\partial L_\epsilon }\,l_\epsilon /l_P^2 \cr &-&i\sum_p \frac{\partial\Gamma}{\partial \phi_p }\,\chi_p /l_P^2 \Big{]}\,,\label{gmea}\eea
where $\bar S = S_{Rc} - i l_P^2 \log \m  + G_N S_{m}$, see \cite{mv}.

We will look for a perturbative solution
$$ \G(L,\f) = S(L,\f) + l_P^2 \G_1 (L,\f) + l_P^4 \G_2 (L,\f) + \cdots \,,$$
and require it to be semiclassical for $L_\e \gg l_P$  and $|\sqrt{G_N}\,\f | \ll 1$. This can be checked on the $E=1$ toy model
$$S(L,\f) =( L^2 + L^4/L_c^2 )\theta(L) + L^2\theta(L)\f^2 (1+  \o^2 L^2 + \lambda\f^2 L^2)\,,$$
where $\theta(L)$ is a homogeneous function of degree zero.

It is not difficult to see that 
$$\G (L,\f) = \G_g (L) + \G_m (L,\f)\,,$$
and 
$$\G_m (L,\f) = V_4 (L) \, U_{eff} (\f)$$
for constant $\f $ where $U_{eff}(0) =0$. Furthermore,
$$\G_g (L) = \G_{pg}(L) + \G_{mg}(L) \,,$$ 
where $\G_{pg}$ is the pure gravity contribution and $\G_{mg}$ is the matter induced contribution. 

In the smooth-manifold approximation one has
$$ \G_{mg}(L) \approx \L_m V_M + \Omega_m (R,K) \,, $$
where $K=2\pi\hbar/L_K$ is the momentum cutoff. One can show that 
$$\O_m = \O_1 l_P^2 + O( l_P^4)$$
and
\bea \O_1 (R,K) &=& a_1 K^2 \int_M d^4 x \sqrt{|g|}\,R \cr
 &+&  \log(K/\o)\, \int_M d^4 x \sqrt{|g|}{\Big [ }a_2 R^2 + a_3 R^{\m\n}R_{\m\n} 
+ a_4 R^{\m\n\r\s} R_{\m\n\r\s}+ a_5\nabla^2 R {\Big ]}\cr
&+& O(1/K^2) \,,\label{krun}\eea
where $R_{\m\n\r\s}$ is the Riemann curvature tensor, see \cite{mv}.

The effective CC will be then given as
$$\L =\L_c +\L_{qg} +\L_m \,,$$
where $\L_{qg}$ is given by (\ref{qgcc}). Note that the matter contribution to CC can be approximated by a sum
\be \L_m \approx \sum_\g v(\g, K) \ee 
where $v(\g,K)$ is a one-particle irreducible vacuum Feynman diagram for the field-theory action $S_m$ in flat spacetime with the  cutoff $K$. One can show that 
\bea \sum_\g v(\g, K) &\approx & l_P^2 \, K^4 {\Big [} c_1 \ln (K^2/\o^2)+ \sum_{n\ge 2}c_n  (\bar\l)^{n-1}(\ln (K^2/\o^2))^{n-2} \cr
 &+& \sum_{n\ge 4} d_n(\bar\l)^{n-1}( K^2/\o^2 )^{n-3} {\Big ]}\,,\label{qftcc}\eea
for $K \gg \omega$, where $\bar\l = l_P^2 \l$, see \cite{mvp}. Therefore one has a highly divergent sum of matter vacuum-energy contributions to the cosmological constant when $K\to\infty$. 

This demonstrates the great difficulty for obtaining the value of the cosmological constant in a perturbative QFT approach to quantum gravity, see \cite{ccrev}. Even if one assumes that there is some non-perturbative value for $\L_m$, one still has the problem of how  to calculate the QG contribution to $\L$, given that GR is a nonrenormalizable QFT. A plausible  assumption is that
$$\L_{qg} + \L_m \approx {1\over l_P^2} \,,$$
so that
\be\L \approx  {1\over l_P^2} + \L_c \,.\label{ftcc}\ee

The problem with (\ref{ftcc}) is that the observed $\L$ has a value $l_P^2 \L \approx 10^{-122}$, so that the free parameter $\L_c$ has to be chosen such that $l_P^2 \L_c$ has to be different from 1 at the 122-nd decimal place, which is an example of the extreme fine-tunning. This extreme fine tunning would not be a problem if the relation (\ref{ftcc}) was an exact result, so that any value of $\L_c$ would be acceptable. However,  (\ref{ftcc}) is not an exact relationship, so that the extreme fine tuning indicates that something is wrong with the assumptions which lead to (\ref{ftcc}).

In the PL formulation of quantum gravity, there is no problem of fine tunning, since we have an exact relationship between $\L$ and the free parameters of the theory.  Namely, one has $\L_{qg} = l_P^2 /L_0^4$ and $\L_m$ is determined by a solution of the EA equation. Therefore  
\be \L_m = V( m^2 , \lambda, l_P^2 ) \,,\ee 
so that
\be\L = \pm {1\over L_c^2} + {l_P^2 \over 2L_0^4} + V(m^2 , \lambda, l_P^2 )\,. \label{efcc}\ee

The free parameters are $L_0$ and  $L_c$. By equating $\L$ with the experimentally observed value, we obtain
\be \l^* = x + y + \l_m \label{pe}\ee
where $\l^* = l_P^2 \L \approx 10^{-122}$, $x = \pm\, l_P^2 / L_c^2$, $y =l_P^4 / 2L_0^4$ and $\l_m = l_P^2 V$. The equation (\ref{pe}) has infinitely many solutions, but we also have to impose the condition for the existence of the semi-classical limit (\ref{pecc}). This gives the restriction 
\be 0 < y < 2|x| \,.\label{scr}\ee 

The value of $\l_m$ is not known, but for any value of $\l_m$ the equation (\ref{pe}) has infinitely many solutions which obey the restriction (\ref{scr}). Note that the solution $x = -\l_m$ and $y = \l^*$, which was proposed in \cite{mv}, will be acceptable if $|\l_m | > \l^* / 2$. This solution is special because it gives a value for $L_0$ which is independent of the value of $\l_m$, which is $L_0 \approx 10^{-5}$m. This is the same value which was obtained in the case of pure PL gravity without the cosmological constant \cite{m}.

Note that  in the case of string theory the cosmological constant can be expressed as a function of the discrete moduli parameters, so that the CC spectrum is discrete \cite{rp}. The CC spectrum in Regge PFQG is continious, so that there is no problem in mathching the observed CC value with the values of the parameters. However, in string theory there are no free parameters, and therefore one has to show that the CC spectrum is sufficiently dense in the interval $[0, 1]$, so that the observed CC value, which is  of the order of $10^{-122}$, can belong to the spectrum. One can argue that this is the case, because there are $10^{500}$ values in the string CC spectrum \cite{pol}. However, this not a proof, but only a plausibility argument.

\section{Conclusions}

Testing the idea that at short distances the structure of the spacetime is given by a PL manifold will require the experimental verification. The distance scale where one could see the PL structure is not necessarilly of the order of Planck length.  It could be 100 or 1000 times smaller than the smallest distance we have probed so far,  which is $10^{-20}\,m$ in the LHC experiments. In those experiments we have not seen any significant deviations from the QFT predictions. The fact that we still see a smooth spacetime at LHC energies may mean that the minimum edge length in the local spacetime triangulation  is several orders of magnitude smaller than $10^{-20}\,m$.

Also note that one can have spatial triangulations\footnote{We assume that after a certain time from the Big Bang the Universe has a topology $\S\times [a,b]$ and the PL structure is given by a causal triangulation (\ref{ct}). We will refer to this situation as the PL structure of the late-time Universe.} such that in the vicinity of our planet we have small edge lengths, smaller than $10^{-20}\,m$,  while in the interstelar space we can have larger, or even much larger edge lengths. Then the propagation of light or $\g$-rays from distant sources could be significantly affected by the PL structure of the spacetime, due to the refraction effects at the boundaries of flat 4-simplices. Basically, the time of arrival of light from a distant source to Earth will depend on the moment of emission since the number of triangles where the spacetime curvature is concentrated and where the light is refracted varies with time. This effect is different from the effect of time-dependent arrival of light due to possible frequency-dependent speed of light, see \cite{grb}. These effects could be observed, given a sufficient precision of the detectors. The same applies to the gravitational waves.

The modification of the exponentially damped four-volume PI measure by a product of inverse powers of the edge lengths given by (\ref{mup}) was introduced in order to achieve the finitenness of the path integral. This modification is of a subleading order in $\log\m$ for large edge lengths with respect to the 4-volume term, so that it did not change the classical limit of the effective action nor affects the cosmological constant analysis. However, for large $N$ and small $L_\e$, the measure modification contribution to the effective action will be negligible in relation to the usual trace-log term. Hence in the spacetime regions where the triangulation can be approximated by a smooth spacetime (the late-time Universe) the physics does not depend on the values of the parameters $l_0$ and $p$.

The $p > \ha$ measure (\ref{mup}) has all the desireable features as the the $p=0$ measure (\ref{plm}), with the added bonus of the possibility to construct the non-perturbative effective action as an exact solution of the EA equation. The exact solutions of the EA equation will be relevant for the early-time universe, when the QG effects are large. 

As far as the CC problem is concerned, the existence of the exact relationship (\ref{efcc}) solves the difficulty of fine tuning. The relation (\ref{efcc}) also distinguishes the Regge PFQG approach from other QG approches, since it allows one to relate the observed CC value with the values of the parameters of the theory.

\end{document}